\documentclass{article}
\usepackage{a4wide}
\usepackage{epsfig}
\usepackage{latexsym}
\def\be{\begin{equation}}
\def\ee{\end{equation}}
\def\bea{\begin{eqnarray}}
\def\eea{\end{eqnarray}}

\title{Out-of-equilibrium quantum fields with conserved charge}

\author{D.~J.~Bedingham \thanks{email:{\tt d.j.bedingham@sussex.ac.uk}}\\
{\small\it Centre for Theoretical Physics, University of Sussex,}\\
{\small\it Falmer, Brighton BN1 9QJ, U.K.} }

\date{\today}

\begin{document}
\maketitle

\begin{abstract}
We study the out-of-equilibrium evolution of an $O(2)$-invariant
scalar field in which a conserved charge is stored. We apply a
loop expansion of the 2-particle irreducible effective action to
3-loop order. Equations of motion are derived which conserve both
total charge and total energy yet allow for the effects of
scattering whereby charge and energy can transfer between modes.
Working in (1+1)-dimensions we solve the equations of motion
numerically for a system knocked out of equilibrium by a sudden
temperature quench. We examine the initial stages of the charge
and energy redistribution. This provides a basis from which we can
understand the formation of Bose-Einstein condensates from first
principles.

\medskip \noindent PACS numbers: 03.75.Nt, 11.30.Fs, 64.90.+b.
\end{abstract}


\section{Introduction}

The formation of a Bose-Einstein condensate involves a significant
proportion of the total charge in a system relocating to occupy
the lowest energy state. This is achieved experimentally using
dilute atomic gases where the conserved charge corresponds to the
total number of atoms and the condensate is formed by reducing the
temperature of the gas using laser and evaporative cooling. We can
understand Bose-Einstein condensation (BEC) by considering the
equilibrium state of a bosonic gas \cite{h&w,kap,ben,dal}, but the
question of precisely how the condensate arises as the gas
responds to a change in its external conditions is a much more
difficult problem (notable examples can be found in
Refs.\cite{stoof1,stoof2,gard}). This process is a very-many-body
problem encompassing the properties of cold, trapped atoms and
importantly, interatomic interactions.

A theoretical understanding of condensate formation has recently
been motivated by the possibility of observing the spontaneous
formation of defects \cite{zurek}. The vortices already observed
in rotating condensates could form spontaneously if the cooling
process were to happen quickly enough. The mechanism by which this
occurs has consequences not only for condensed matter physics but
also for cosmology where it is predicted that similar defects may
have formed spontaneously in the early universe \cite{kibble}.
Experimental production of spontaneous defects in atomic gases
would effectively allow us to test cosmological theories in the
laboratory.

In this article we shall consider the initial stages of the
dynamical process by which charge and energy is redistributed
between modes following a sudden drop in the ambient temperature.
Recent progress has been made towards understanding such
equilibration processes of quantum fields in
Refs.\cite{berges1,berges2,aart,aart2,bett,hein}. In particular,
the method of Refs.\cite{berges1,berges2,aart2} uses the loop
expansion for the 2-particle irreducible (2PI) effective action
where thermalization is observed at 3-loop order \cite{berges1}.
Here we apply this method to a system containing a conserved
finite charge.

In the next section we outline our method for a relativistic
scalar field theory invariant under $O(2)$-transformations (the
non-relativistic theory emerges for low temperatures as is
demonstrated in~\cite{tim1}). This is one of the simplest models
to have a conserved Noether charge, representing some conserved
quantum number. We derive equations of motion describing the
evolution of the Green functions of the theory. In section
\ref{sec:3} we consider the initial conditions needed to fully
describe the system. For convenience we choose an initial
equilibrium state. In section \ref{sec:4} we solve the equations
of motion numerically. Due to computational constraints we work in
one space and one time dimension. Though there is no phase
transition in (1+1)-dimensions, trapped finite-size gases are
known to exhibit interesting phenomena associated with a
macroscopic occupation of the lowest energy state \cite{dal,toms}.
Here we examine the initial stages of the charge and energy
redistribution which must form the basis by which BEC occurs.


\section{Dynamical equations}
\label{sec:2}

To describe out-of-equilibrium behavior we use the
Schwinger-Keldysh technique \cite{sch,kel,land}. Given an initial
distribution of states with density operator $\hat{\rho}$ at time
$t_i$, the expectation of an operator $\hat{\cal O}$ at some later
time $t$ is given in the Heisenberg picture by
    \be
    \langle\hat{\cal O}(t)\rangle = {\rm Tr}\left\{\hat{\rho}\hat{\cal
    O}(t)\right\}.
    \label{eq:OP}
    \ee
In the path-integral formulation, the time contour $C$ begins at
time $t_i$, evolves forward to time $t$, then returns to $t_i$.

We are interested in considering the response of a scalar field to
a rapid drop in temperature. Assuming that the system is in
equilibrium prior to the temperature quench, we may take the
original equilibrium distribution as our initial distribution of
states. Following the quench, as the system evolves with lower
energy, it is forced to seek a new equilibrium.

Since our initial state has non-zero charge, the grand canonical
ensemble provides the obvious choice of density operator. Given a
Hamiltonian $\hat{H}$, charge $\hat{Q}$, initial temperature
$T=1/\beta$ and chemical potential $\mu$, this is given by
   \be
    \hat{\rho} = \exp\left\{-\beta\left(\hat{H}-\mu
    \hat{Q}\right)\right\}.
   \ee
We are free to adjust the effective parameters of $\hat{H}$ in
order to fix our initial distribution. The initial conditions are
considered in more detail in section \ref{sec:3}.

The Hamiltonian density and charge operators for the
self-interacting $O(2)$-invariant scalar theory are
    \bea
        {\cal H}&=&
        \frac{1}{2}\pi_a\pi_a+\frac{1}{2}\nabla\phi_a\nabla\phi_a
        +\frac{1}{2}m^2\phi_a\phi_a+\frac{\lambda}{4!}\left(\phi_a\phi_a\right)^2,
        \nonumber\\
        Q&=&\int {\rm d} x
        \left(\pi_1\phi_2-\pi_2\phi_1\right),
    \eea
where $a=1,2$ and repeated indices are summed over in the usual
way. We proceed by writing out the Schwinger-Dyson equation for
the 2-point Green function of this field \cite{land} (we suppress
field indices initially)
    \be
    \label{eq:S-D}
        (\Box_x+m^2)G_2(xx_1)+\frac{\lambda}{6}G_4(xxxx_1)=-i
        \delta(x-x_1).
    \ee
Diagrammatically, the 2-point Green function can be represented by
a line as follows
    \bea
        G_2(x_1x_2)=
    \begin{picture}(8,4)
    \put(2,0.5){\makebox(0,0){$x_1$}}
    \put(13,0.5){\makebox(0,0){$x_2$}}
    \put(4,0.8){\line(1,0){6}}
    \put(4,0.8){\circle*{1}}\put(10,0.8){\circle*{1}}
    \end{picture}
    \eea
The 4-point Green function can then be written as a coupling
expansion in terms of the 2-point Green function
    \bea
        G_4(x_1x_2x_3x_4)=
    \begin{picture}(16,4)
    \put(2,4.5){\makebox(0,0){$x_1$}}
    \put(13,4.5){\makebox(0,0){$x_2$}}
    \put(2,-2.5){\makebox(0,0){$x_3$}}
    \put(13,-2.5){\makebox(0,0){$x_4$}}
    \put(4,4){\line(1,0){6}}
    \put(4,4){\circle*{1}}\put(10,4){\circle*{1}}
    \put(4,-2){\line(1,0){6}}
    \put(4,-2){\circle*{1}}\put(10,-2){\circle*{1}}
    \end{picture}
    +
    \begin{picture}(16,4)
    \put(2,4.5){\makebox(0,0){$x_1$}}
    \put(13,4.5){\makebox(0,0){$x_3$}}
    \put(2,-2.5){\makebox(0,0){$x_2$}}
    \put(13,-2.5){\makebox(0,0){$x_4$}}
    \put(4,4){\line(1,0){6}}
    \put(4,4){\circle*{1}}\put(10,4){\circle*{1}}
    \put(4,-2){\line(1,0){6}}
    \put(4,-2){\circle*{1}}\put(10,-2){\circle*{1}}
    \end{picture}
    +
    \begin{picture}(16,4)
    \put(2,4.5){\makebox(0,0){$x_1$}}
    \put(13,4.5){\makebox(0,0){$x_4$}}
    \put(2,-2.5){\makebox(0,0){$x_2$}}
    \put(13,-2.5){\makebox(0,0){$x_3$}}
    \put(4,4){\line(1,0){6}}
    \put(4,4){\circle*{1}}\put(10,4){\circle*{1}}
    \put(4,-2){\line(1,0){6}}
    \put(4,-2){\circle*{1}}\put(10,-2){\circle*{1}}
    \end{picture}
    -i\lambda\int {\rm d} y\;
    \begin{picture}(16,4)
    \put(2,4.5){\makebox(0,0){$x_1$}}
    \put(13,4.5){\makebox(0,0){$x_2$}}
    \put(2,-2.5){\makebox(0,0){$x_3$}}
    \put(13,-2.5){\makebox(0,0){$x_4$}}
    \put(9.5,1){\makebox(0,0){$y$}}
    \put(7,1){\circle*{1}}
    \put(4,4){\line(1,-1){6}}
    \put(4,4){\circle*{1}}\put(10,4){\circle*{1}}
    \put(4,-2){\line(1,1){6}}
    \put(4,-2){\circle*{1}}\put(10,-2){\circle*{1}}
    \end{picture}+\cdots.
    \label{eq:g4}
    \eea
Setting $x_2=x_2=x_3=x$ we have
    \bea
    G_4(xxxx_1)= 3
    \begin{picture}(16,4)
    \put(4,1){\line(1,0){6}}
    \put(4,3){\circle{4}}
    \put(4,1){\circle*{1}}\put(10,1){\circle*{1}}
    \put(2,0.5){\makebox(0,0){$x$}}\put(13,0.5){\makebox(0,0){$x_1$}}
    \end{picture}
    -i\lambda\int {\rm d} y\;
    \begin{picture}(22,4)
    \put(4,1){\line(1,0){12}}
    \put(7,1){\circle{6}}
    \put(4,1){\circle*{1}}\put(10,1){\circle*{1}}\put(16,1){\circle*{1}}
    \put(2,0.5){\makebox(0,0){$x$}}\put(11.5,-0.5){\makebox(0,0){$y$}}
    \put(19,0.5){\makebox(0,0){$x_1$}}
    \end{picture}+\cdots.
    \label{eq:g4xxx}
    \eea
The first term on the right-hand side results from disconnected
2-point Green functions.  Truncating the series at this term
constitutes a `mean-field' linearization of Eq.(\ref{eq:S-D}),
known as the Hartree approximation. Since we wish to incorporate
scattering of particles with the movement of charge and energy
between modes, it is crucial that we keep at least the ${\cal
O}(\lambda)$ term in the series expansion. It seems clear when we
look at Eq.(\ref{eq:g4}) that the last term is necessary if we
wish to see scattering effects.

Replacing field indices and neglecting the point-number label (all
Green functions under consideration from now on are 2-point) we
may rewrite the Schwinger-Dyson equation as
    \be
    \label{eq:sd2}
        (\Box_x\delta_{ac}+M_{ac}^2(x))G_{cb}(x,y)
        +i\int_{C}{\rm d}z
        \Sigma_{ac}(x,z)G_{cb}(z,y)=-i\delta_{ab}\delta_{C}(x-y),
    \ee
where $M$ is the effective mass and $\Sigma$ is the non-local
self-energy. Using the expansion of Eq.(\ref{eq:g4xxx}) and
reinterpreting the algebraic form of the diagrams gives the
effective mass as
    \be
        M_{ab}^2(x)=m^2\delta_{ab}+\frac{\lambda}{6}\left[G_{cc}(x,x)\delta_{ab}+2G_{ab}(x,x)\right],
    \ee
and the non-local self-energy as
    \be
        \Sigma_{ab}(x,y)=
        -2\left(\frac{\lambda}{6}\right)^2\left[G_{cd}(x,y)G_{cd}(x,y)G_{ab}(x,y)
        +2G_{ac}(x,y)G_{db}(x,y)G_{dc}(x,y)\right].
        \label{eq:siggy}
    \ee
We began with the Schwinger-Dyson equation and performed a
coupling expansion of the 4-point Green function. The above result
is equivalently obtained by starting with the 2PI effective action
and truncating at 3-loop order as is shown in
Refs.\cite{berges2,aart2} (e.g. our Eq.(\ref{eq:sd2}) corresponds
to Eq.(13) of Ref.\cite{aart2}). In each of these references, a
$1/N$ expansion is used in order to derive $M$ and $\Sigma$.
Although in principle our coupling expansion can be derived by
truncating the $1/N$ expansion at ${\cal O}(\lambda^2)$, this
would require working to NNLO in order to encounter the ${\cal
O}(1/N^2)$ term on the right hand side of Eq.(\ref{eq:siggy}). The
weak-coupling limit of the NLO approximation is briefly considered
in Ref.\cite{aart2}.

The problem we face is to solve Eq.(\ref{eq:sd2}) for some given
set of initial conditions. The 2-point Green functions represent a
time-contour ordered product of field operators
    \be
        G_{ab}(x,y)=\langle T_{C} \phi_a(x)
        \phi_b(y)\rangle.
    \ee
The usual procedure for solving the Schwinger-Dyson equation is to
decompose the Green functions as \cite{land}
    \be
        G_{ab}(x,y)=\theta_{C}(x_0-y_0)G_{ab}^>(x,y)+\theta_{C}(y_0-x_0)G_{ab}^<(x,y)
    \ee
such that
    \bea
        G_{ab}^>(x,y)&=&\langle \phi_a(x)\phi_b(y)\rangle,\nonumber\\
        G_{ab}^<(x,y)&=&\langle \phi_b(y)\phi_a(x)\rangle.
    \eea
Now following the method outlined in Refs.\cite{berges2,aart2} we
take real and imaginary parts
    \bea
        F_{ab}(x,y)&=&\frac{1}{2}\left[G_{ab}^>(x,y)+G_{ab}^<(x,y)\right]={\rm Re} G_{ab}^>(x,y),\nonumber\\
        \rho_{ab}(x,y)&=&i\left[G_{ab}^>(x,y)-G_{ab}^<(x,y)\right]=-2{\rm Im}
        G_{ab}^>(x,y).
        \label{eq:gfd}
    \eea
The function $\rho$ is the spectral function and $F$ is the
symmetric propagator.

Similarly, we perform a decomposition of the self-energy
    \be
        \Sigma_{ab}(x,y)=\theta_{C}(x_0-y_0)\Sigma_{ab}^>(x,y)
        +\theta_{C}(y_0-x_0)\Sigma_{ab}^<(x,y),
    \ee
and again take real and imaginary parts
    \bea
        \Sigma^F_{ab}(x,y)&=&\frac{1}{2}\left[\Sigma_{ab}^>(x,y)+\Sigma_{ab}^<(x,y)\right],\nonumber\\
        \Sigma^{\rho}_{ab}(x,y)&=&i\left[\Sigma_{ab}^>(x,y)-\Sigma_{ab}^<(x,y)\right].
    \eea

We can now reexpress the Schwinger-Dyson equation in terms of $F$
and $\rho$ (cf. Eqs.(70) and (71) in Ref.\cite{aart2})
    \bea
        (\Box_x\delta_{ac}+M_{ac}^2)F_{cb}(x,y)&=&-\int {\rm d}{\bf
        z}\left\{\int_0^{x_0}{\rm
        d}z_0\Sigma_{ac}^{\rho}(x,z)F_{cb}(z,y)-\int_0^{y_0}{\rm
        d}z_0\Sigma_{ac}^{F}(x,z)\rho_{cb}(z,y)\right\},\nonumber\\
        (\Box_x\delta_{ac}+M_{ac}^2)\rho_{cb}(x,y)&=&-\int {\rm d}{\bf
        z}\left\{\int_{y_0}^{x_0}{\rm
        d}z_0\Sigma_{ac}^{\rho}(x,z)\rho_{cb}(z,y)\right\},
    \eea
where now
    \be
        M_{ab}^2(x)=m^2\delta_{ab}+\frac{\lambda}{6}\left[F_{cc}(x,x)\delta_{ab}+2F_{ab}(x,x)\right]
    \ee
and
    \bea
        \Sigma_{ab}^F=-2\left(\frac{\lambda}{6}\right)^2&&\left[
        \left(F_{cd}F_{cd}-\frac{1}{4}\rho_{cd}\rho_{cd}\right)F_{ab}
        -\frac{1}{2}F_{cd}\rho_{cd}\rho_{ab}\right.\nonumber\\
        &&\left.+2\left(F_{ac}F_{db}F_{dc}
        -\frac{1}{4}F_{ac}\rho_{db}\rho_{dc}
        -\frac{1}{4}\rho_{ac} F_{db}\rho_{dc}
        -\frac{1}{4}\rho_{ac}\rho_{db} F_{dc}\right)
        \right],\nonumber\\
        \Sigma_{ab}^{\rho}=-2\left(\frac{\lambda}{6}\right)^2&&\left[
        \left(F_{cd}F_{cd}-\frac{1}{4}\rho_{cd}\rho_{cd}\right)\rho_{ab}
        +2F_{cd}\rho_{cd}F_{ab}\right.\nonumber\\
        &&\left.+2\left(\rho_{ac}F_{db}F_{dc}
        +F_{ac}\rho_{db}F_{dc}
        +F_{ac} F_{db}\rho_{dc}
        -\frac{1}{4}\rho_{ac}\rho_{db} \rho_{dc}\right)
        \right].
    \eea

Before attempting to solve these equations, we express them in
momentum space. In general, given that $\int_{\rm p}=\int{\rm
d}^dp/(2\pi)^d$, we can write
    \be
        G(t_1,t_2;{\bf x-y})=\int_{\rm p}e^{i{\bf
        p(x-y)}}G(t_1,t_2;{\bf p})
    \ee
due to spatial translation invariance. The self-energy terms
$\Sigma^F$ and $\Sigma^{\rho}$ take the form given above but
factors in each term have momentum values $\bf p_1$, $\bf p_2$ and
$({\bf p-p_1-p_2})$ respectively. Integrations are performed over
$\bf p_1$ and $\bf p_2$ and the self-energy carries a momentum
label $\bf p$.

The dynamical equations become
   \bea
        \left[(\partial^2_{t_1}+{\bf
        p}^2)\delta_{ac}+M_{ac}^2(t_1)\right]
        F_{cb}(t_1,t_2;{\bf p})&=&-\left\{\int_0^{t_1}{\rm
        d}t'\Sigma_{ac}^{\rho}(t_1,t';{\bf p})F_{cb}(t',t_2;{\bf p})\right.\nonumber\\
        &&\;\;\;\;\;\;\;\;\left.-\int_0^{t_2}{\rm
        d}t'\Sigma_{ac}^{F}(t_1,t';{\bf p})\rho_{cb}(t',t_2;{\bf p})\right\},\nonumber\\
        \left[(\partial^2_{t_1}+{\bf
        p}^2)\delta_{ac}+M_{ac}^2(t_1)\right]
        \rho_{cb}(t_1,t_2;{\bf p})&=&-\left\{\int_{t_2}^{t_1}{\rm
        d}t'\Sigma_{ac}^{\rho}(t_1,t';{\bf p})\rho_{cb}(t',t_2;{\bf
        p})\right\},
        \label{eq:cde}
    \eea
where the effective mass squared, expressed in the momentum basis
is
    \be
        M_{ab}^2(t)=m^2\delta_{ab}+\frac{\lambda}{6}\left[\int_{\rm p}F_{cc}(t,t;{\bf p})\delta_{ab}
        +2\int_{\rm p}F_{ab}(t,t;{\bf p})\right].
        \label{eq:mass}
    \ee
It remains to solve these equations for some given set of initial
conditions. Since our equations involve a weak-coupling expansion
we should be careful to note that they are likely to give
inaccurate answers should the coupling corrections become large.
For this reason we should not stray too far from equilibrium.

Finally, by taking expectations of the Hamiltonian and charge
operators and substituting from Eq.(\ref{eq:g4xxx}) for the
4-point Green function, we find the overall energy is given by
    \bea
        E&=&\int_{\rm p}E_{\rm p}(t)\nonumber\\
        &=& \frac{1}{2}\int_{\rm p}\left\{
            \left[\partial_{t}\partial_{t'}F_{aa}(t,t';{\bf p})\right]_{t=t'=0}
            +\left({\bf
            p}^2+m^2+\frac{\lambda}{24}\int_{\rm p'}F_{cc}(t,t;{\bf
            p'})\right)F_{aa}(t,t;{\bf p})
            \right.\nonumber\\ && \;\;\;\;\;\;\;\;
            +\left(\frac{\lambda}{12}\int_{\rm p'}F_{ac}(t,t;{\bf p'})\right)F_{ca}(t,t;{\bf p})
            \nonumber\\ && \;\;\;\;\;\;\;\;
            +\left.\frac{1}{2}\int_0^{t}{\rm
             d}t'\left(\Sigma_{ac}^{\rho}(t,t';{\bf p})F_{ca}(t',t;{\bf p})
             -\Sigma_{ac}^{F}(t,t';{\bf
             p})\rho_{ca}(t',t;{\bf
            p})\right)\right\}
            \label{eq:ener}
        \eea
and the overall charge is given by
    \be
    Q= \int_{\rm p}Q_{\rm p}(t) =  \int_{\rm p} \left[\partial_{t}F_{12}(t,t';{\bf p})
        -\partial_{t}F_{21}(t,t';{\bf p})\right]_{t=t'}
        \label{eq:char}.
    \ee
Each of these quantities can be shown both analytically and
numerically to be conserved by the dynamical equations. However,
charge and energy are indeed capable of exchange between modes.


\section{Initial Conditions}
\label{sec:3}

The Schwinger-Dyson equations embody not only Heisenbergs equation
of motion for the field but also the equal time commutation
relations (ETCR). It should come as no surprise that the ETCR are
conserved by our dynamics so long as they are enforced by the
initial conditions. The ETCR for our model are
    \bea
        \left[\phi_a({\bf x},t),\pi_b({\bf
        y},t)\right]&=&i\delta_{ab}\delta({\bf x}- {\bf
        y}),\nonumber\\
        \left[\phi_a({\bf x},t),\phi_b({\bf
        y},t)\right]=\left[\pi_a({\bf x},t),\pi_b({\bf
        y},t)\right]&=&0.
    \eea
These translate into statements about the spectral function
    \bea
        \left[\partial_{t}\rho_{ab}(t,t';{\bf p})\right]_{t=t'}
            &=&\delta_{ab},\nonumber\\
        \rho_{ab}(t,t;{\bf p})=
        \left[\partial_{t}\partial_{t'}\rho_{ab}(t,t';{\bf p})\right]_{t=t'}
             &=&0.
    \eea
The ETCR are not sufficient to constrain the initial conditions.
Other symmetry requirements which derive from the spectral
function and symmetric propagator definitions Eq.(\ref{eq:gfd})
must also be satisfied
    \bea
        F_{ab}(t_1,t_2;{\bf p})&=&F_{ba}(t_2,t_1;{\bf p}),\nonumber\\
        \rho_{ab}(t_1,t_2;{\bf p})&=&-\rho_{ba}(t_2,t_1;{\bf p}).
    \eea
(It can be checked that if these conditions are satisfied
initially, then the dynamical equations preserve them for all
later times.) Having satisfied the quantum commutation relations
and other symmetry requirements, it remains to chose the initial
conditions for these unspecified functions. Here we shall choose
free field equilibrium values for some given temperature
$T=1/\beta$, and chemical potential $\mu$. Setting the initial
time $t_i=0$ we have
    \bea
        F_{12}(0,0;{\bf p})=F_{21}(0,0;{\bf p})&=& 0,\nonumber\\
        F_{11}(0,0;{\bf p})=F_{22}(0,0;{\bf p})&=&
        \frac{1}{2\omega}\left[\frac{1}{e^{\beta(\omega-\mu)}-1}
            +\frac{1}{e^{\beta(\omega+\mu)}-1}+1\right],\nonumber\\
        \left[\partial_{t}F_{11}(t,0;{\bf p})\right]_{t=0}=
        \left[\partial_{t}F_{22}(t,0;{\bf p})\right]_{t=0}&=&0,\nonumber\\
        \left[\partial_{t}F_{12}(t,0;{\bf p})\right]_{t=0}=
        -\left[\partial_{t}F_{21}(t,0;{\bf p})\right]_{t=0}&=&
        \frac{1}{2}\left[\frac{1}{e^{\beta(\omega-\mu)}-1}
            -\frac{1}{e^{\beta(\omega+\mu)}-1}\right],  \nonumber\\
        \left[\partial_{t}\partial_{t'}F_{ab}(t,t';{\bf p})\right]_{t=t'=0}
        &=&\omega^2 F_{ab}(0,0;{\bf p}),
    \eea
where $\omega=\sqrt{{\bf p}^2+M^2_{0}}$, and $M^2_0$ can be viewed
as the effective mass squared prior to the temperature quench.

Referring to Eq.(\ref{eq:char}), the overall charge is given by
    \be
     Q=\int_{\rm p} \left[\frac{1}{e^{\beta(\omega-\mu)}-1}
            -\frac{1}{e^{\beta(\omega+\mu)}-1} \right].
        \label{eq:QQQ}
   \ee
This value is conserved throughout the subsequent evolution of
Green functions.


\section{Numerical results}
\label{sec:4}

In this section we present our numerical solutions to Eqs.
(\ref{eq:cde}).

Eqs. (\ref{eq:cde}) are integro-differential equations for
$F_{ab}$ and $\rho_{ab}$, each of which are functions of $t_1$,
$t_2$ and ${\bf p}$. The equations involve coefficients which are
themselves dependent on $F_{ab}(t,t';{\bf p})$ and
$\rho_{ab}(t,t';{\bf p})$ for $t$ and $t'$ up to and including
$t_1$ and $t_2$, and for all values of ${\bf p}$. The right hand
side of Eqs (\ref{eq:cde}) involve two integrations over internal
momenta and one time integration. In addition these integrations
must be performed for all external values of $t_1$, $t_2$ and
${\bf p}$. In light of this it would be computationally unfeasible
to work in any greater than (1+1)-dimensions. Also, the range of
the time integrations becomes greater as $t_1$ and $t_2$ become
greater and contrives to slow the calculations down at later
times. This limits us to considering only early times.

The (1+1)-dimensional continuum is replaced by a spatially
periodic lattice. Integrations are performed by converting into
discrete sums
    \be
    \int\frac{{\rm d}{\bf p}}{2\pi}
    \rightarrow\frac{1}{Na}\sum_{n=1}^{N}.
    \ee
$N$ is the number of lattice points, $a$ is the spatial lattice
spacing, and $Na=L$ is the length of the spatial dimension .
Differential equations are solved by a basic leapfrog method. The
time step $h$ in units of $1/M_{\rm INIT}$ is chosen to be 0.1
where $M_{\rm INIT}$ is the initial value of the effective mass
following the quench. A corresponding choice of $a=0.4/M_{\rm
INIT}$ gives stability with $h/a=0.25$. Choosing $N=100$ we have a
lattice size $L=40$ in units of $1/M_{\rm INIT}$. With periodic
boundary conditions the momenta are given by
    \be
    {\bf p}^2\rightarrow -\frac{2}{a^2}\left(\cos\frac{2\pi
    n}{N}-1\right).
    \ee

Although these elementary techniques may provide a relatively poor
approximation to the continuum, they are fast and provide stable
results. Also, the conservation of energy, charge and other
symmetries are transparent upon considering a single iteration of
the dynamical process.

Other parameters are chosen in units of $M_{\rm INIT}$ as $T=10$,
$\mu=0.1$ and $M_0=4$. These correspond to a total energy of
172.92 and a total charge of 1.84. These values have no particular
significance, they are chosen simply to represent the general
behavior. Larger values of $M_0$ provide a greater impulse,
knocking the system further from equilibrium. Larger values of $T$
reduce the impact of changing the mass parameter on the charge and
energy distributions. The chemical potential must be less than
$M_0$. For (3+1)-dimensional systems, the closer $\mu$ is to
$M_0$, the closer the system is to the critical point. The
coupling $\lambda=0.5$ is small in order to make sense of a
coupling expansion yet large enough to see appreciable scattering
effects relatively quickly.

Fig.\ref{fig:1} shows plots with the effective mass squared
$M^2(t)$ (Eq.({\ref{eq:mass})), the energy in the lowest mode
$E_0(t)$ (see Eq.(\ref{eq:ener})), and the charge in the lowest
mode $Q_0(t)$ (see Eq.(\ref{eq:char})). The dashed line is the
Hartree or mean-field approximation for comparison. In the Hartree
approximation only terms of order $\lambda$ are kept in the
dynamical equations. Solid lines represent solutions to
Eqs.(\ref{eq:cde}). An important feature to observe is that the
charge in the lowest energy state is constant in the Hartree
approximation but is free to change at next order in $\lambda$
where scattering effects are incorporated. Oscillations in
$M^2(t)$ and $E_0(t)$ are damped out more effectively in the
higher order calculation.

\begin{figure}[t]
\epsfxsize=18pc
\[\epsfbox{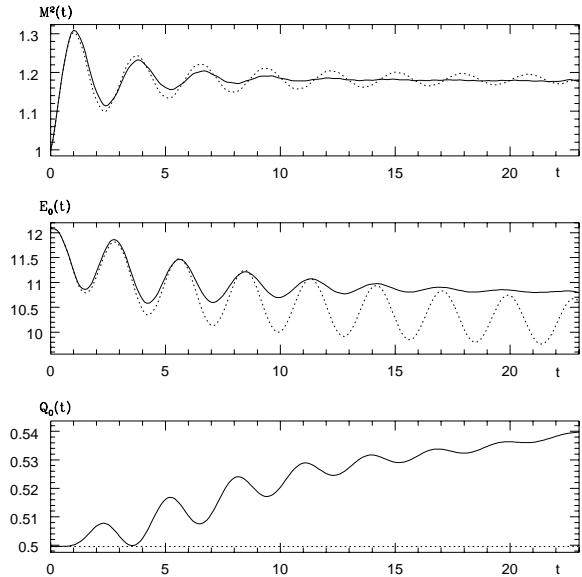}\]
\caption{\small Evolution of effective mass squared, energy in the
lowest mode, and charge in the lowest mode. The dashed line is the
Hartree approximation, the solid line is the current
approximation. } \label{fig:1}
\end{figure}

\begin{figure}[h]
\epsfxsize=18pc
\[\epsfbox{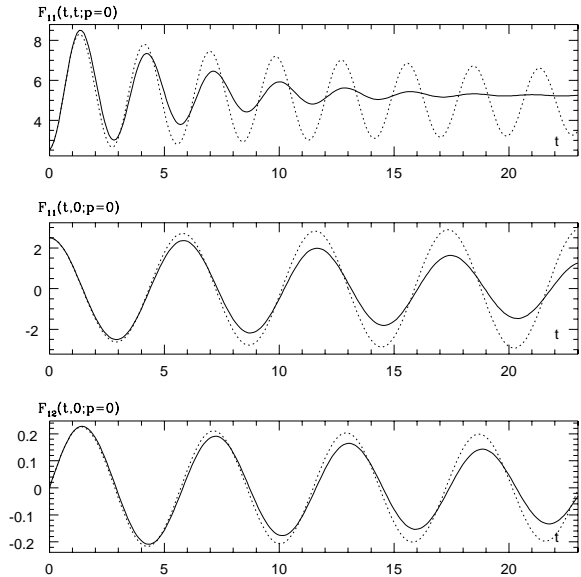}\]
\caption{\small Evolution of the symmetric propagator. The dashed
line is the Hartree approximation, the solid line is the current
approximation.} \label{fig:2}
\end{figure}

Fig.\ref{fig:2} shows the evolution of the symmetric propagator
with time. For $F_{11}(t,t;0) (=F_{22}(t,t;0))$ we observe a much
more effective damping of oscillations than in the Hartree
approximation. We also observe an oscillatory decrease in
correlations between the field at time $t$ and the field at time
$0$. This indicates that the system is moving towards some final
outcome which is independent of the initial conditions and
dependent only on the conserved quantities which characterize the
system. It should be noted that the equations of motion may be
reversed in time and initial conditions can always be recovered. A
fixed point equilibrium solution in the future can in principle be
approached arbitrarily closely but never reached \cite{berges2}.
There is zero equal time correlation between fields $\phi_1$ and
$\phi_2$.

Fig.\ref{fig:3a} shows the distribution of charge in the different
momentum modes (Eq.(\ref{eq:char})). The upper figure indicates
the initial charge distribution (see Eq.(\ref{eq:QQQ})). The lower
momentum modes store a greater amount of charge. As the system
evolves we see that charge moves out-of the region $p > 1$ and
into the region $p < 1$. The change in charge at each momentum
value is shown at times 0, 10, and 20. This process fits our
intuition where we would expect that charge moves into the lower
energy states as the temperature decreases.

\begin{figure}[t]
\epsfxsize=18pc
\[\epsfbox{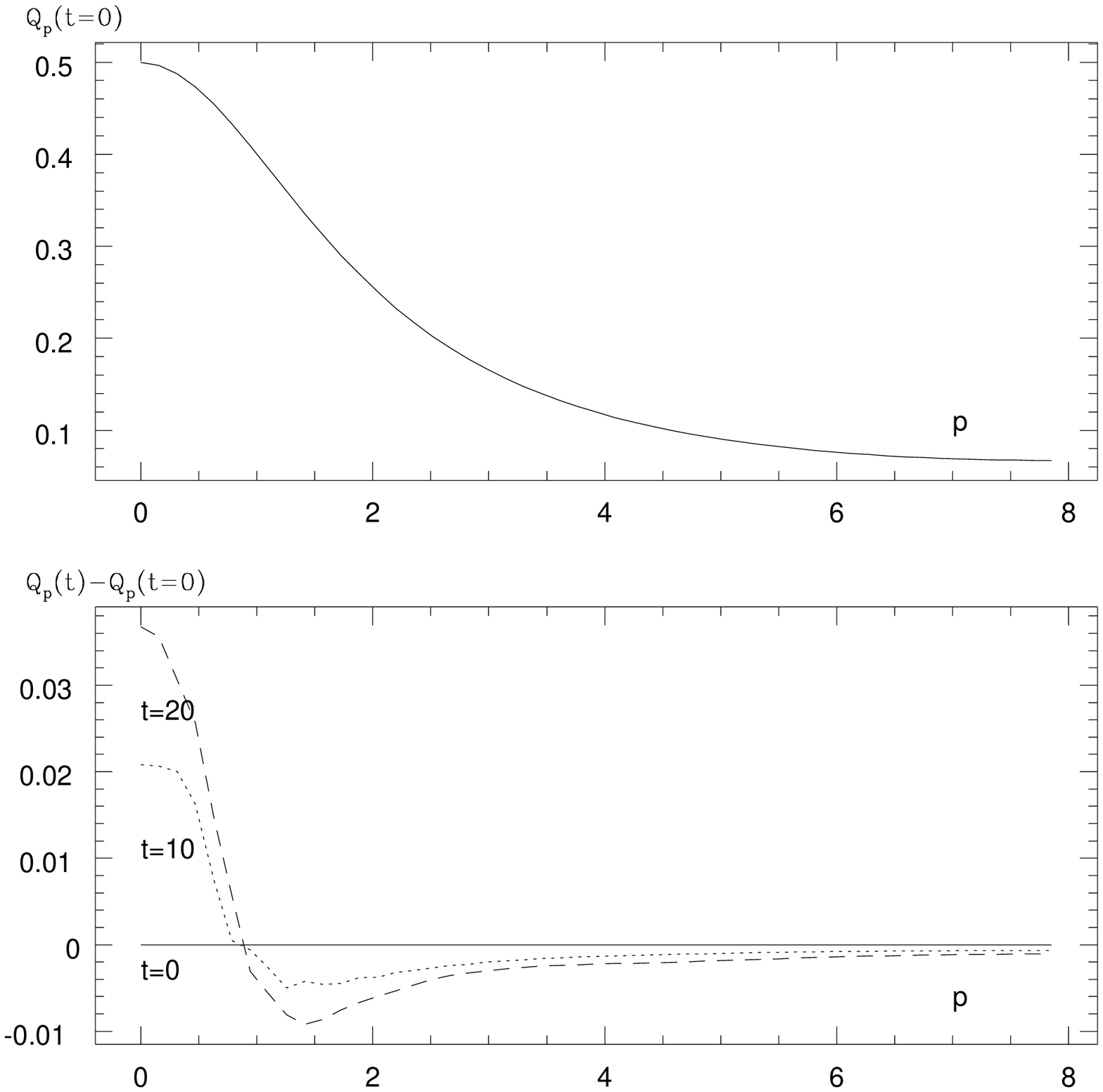}\]
\caption{\small Charge distributions at a series of times.}
\label{fig:3a}
\end{figure}

\begin{figure}[h]
\epsfxsize=18pc
\[\epsfbox{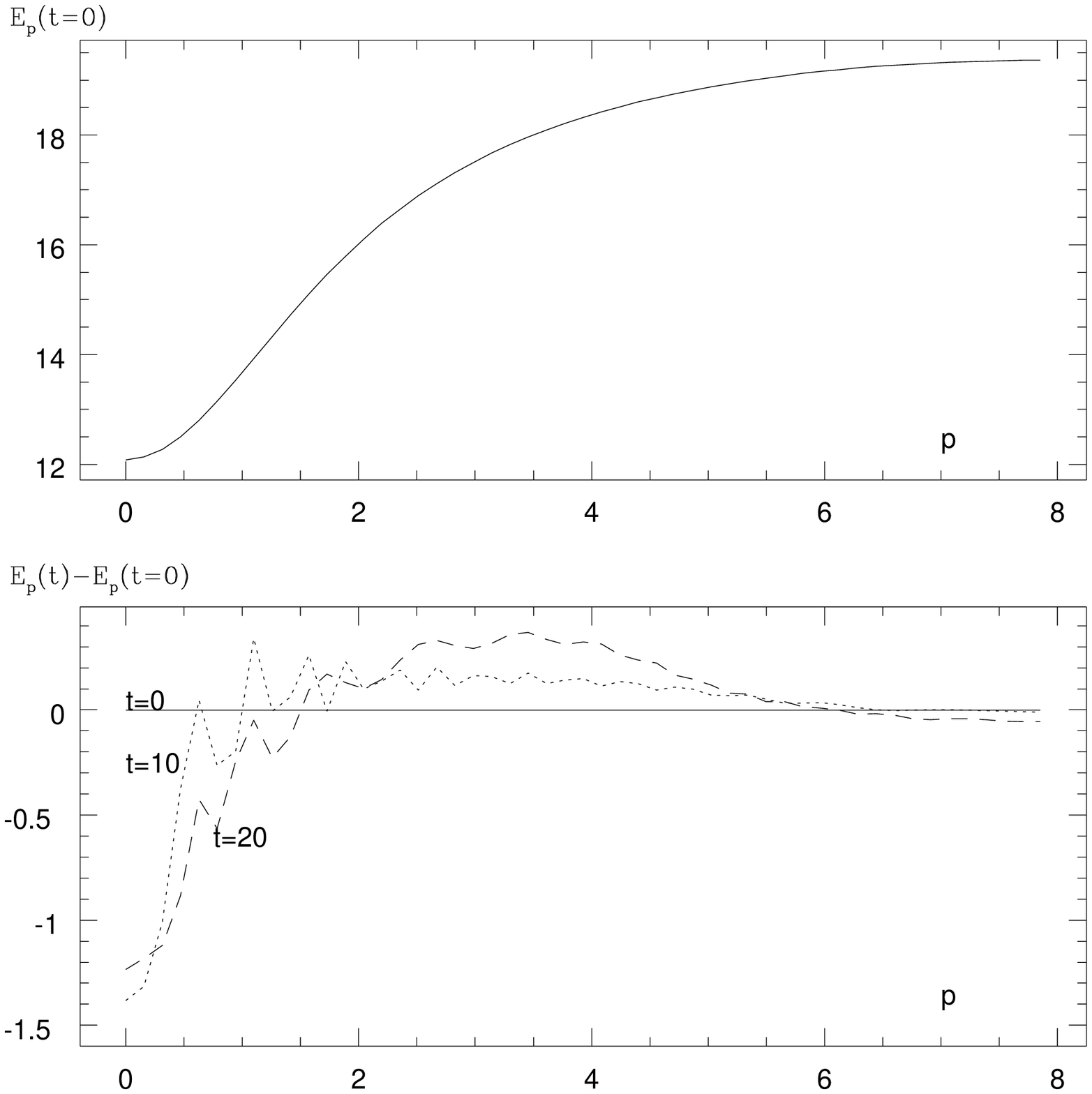}\]
\caption{\small Energy distributions at a series of times.}
\label{fig:3b}
\end{figure}

The energy distribution of Eq.(\ref{eq:ener}) is given in
Fig.\ref{fig:3b}. The upper figure gives the initial energy
distribution. In the continuum we would expect the energy to
increase quadratically with $p$ and without bound. Our lattice
approximation flattens out the energy spectrum at the upper limit
of momentum whilst the spectrum is approximately quadratic at low
momenta. Following the quench we see a movement of the energy from
the region $p < 1$ into the region $p > 1$. The change in the
energy distribution becomes less oscillatory as time increases.
This is a reflection of the fact that oscillations in both the
effective mass and the equal time propagator are damped out as
time increases.

Fig.\ref{fig:4} displays the increase in spatial correlations as a
result of the drop in temperature. The correlation function
    \be
        F_{11}(t,t;{\bf x})=\int_{\rm p}e^{i{\bf
        p x}}F_{11}(t,t;{\bf p})
    \ee
shows a growth in amplitude and a growth in the size of the region
in which the field is correlated. These regions can be interpreted
as domains in which the field has strong correlations \cite{boy}.
The rate at which these domains grow has consequences for the
formation of defects in higher dimensional systems undergoing a
symmetry breaking phase transition \cite{zurek2}. In our
simulations it is found that the correlation function quickly
grows to a new stable function which then remains approximately
constant.

\begin{figure}[t]
\epsfxsize=18pc
\[\epsfbox{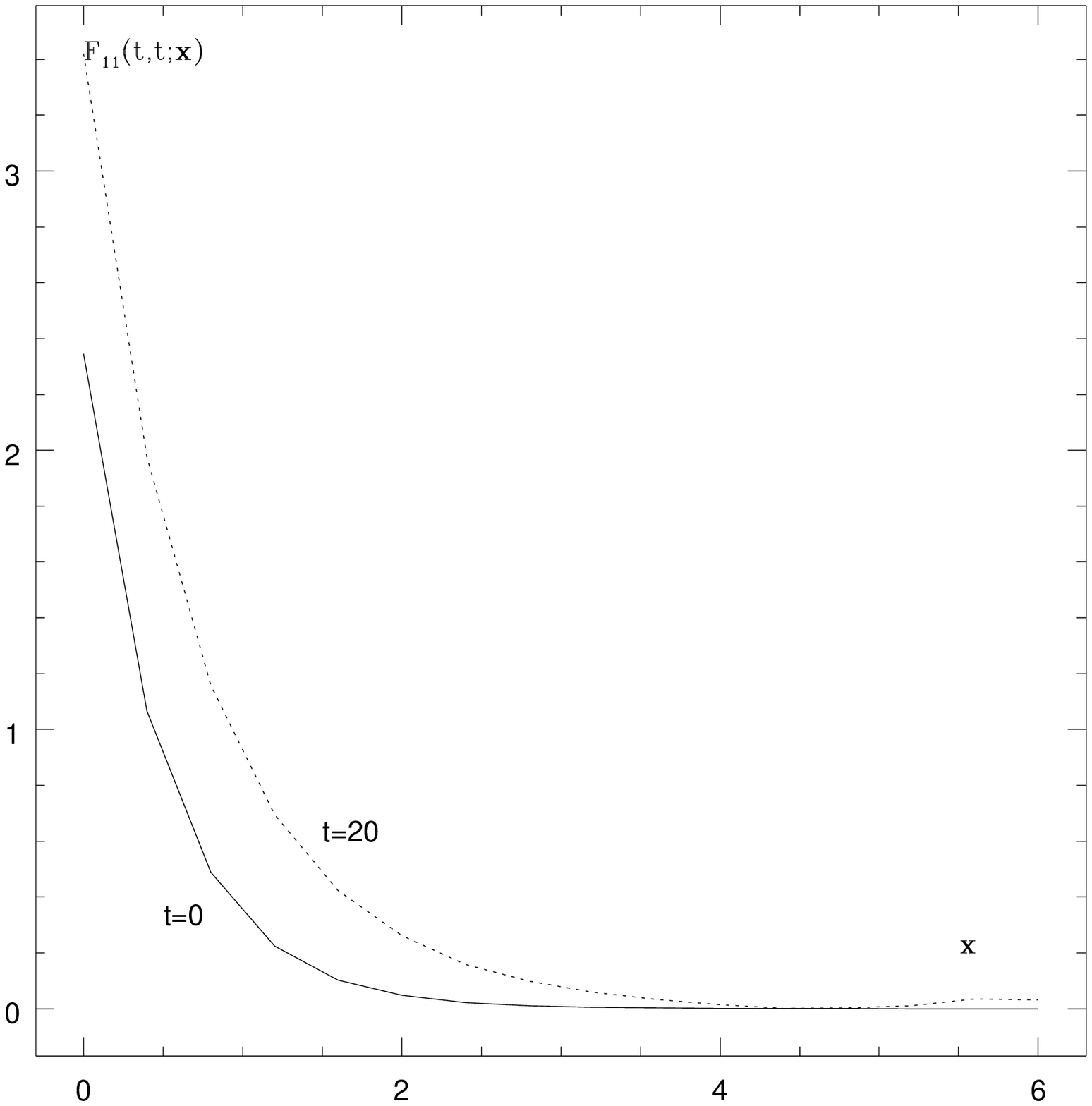}\]
\caption{\small Changes in the spatial correlation function with
time.} \label{fig:4}
\end{figure}

In addition, in our numerics we observe conservation of total
charge, conservation of total energy, and conservation of
commutation relations.


\section{Conclusions}

We have considered the quantum field dynamics of an
$O(2)$-invariant scalar field theory with a conserved charge. We
have argued that in deriving our equations of motion, it is
crucial that we work to at least 3-loop order in the 2-particle
irreducible effective action. This is the lowest order at which
the effects of scattering are included, allowing for the movement
of charge between modes. In lower order calculations, the charge
in each mode is artificially confined to that mode. Since BEC is
essentially characterized by the charge distribution, in order to
consider the formation process, the charge in the system must be
able to redistribute in response to some external stimulus. More
generally, to understand any equilibration process involving
finite conserved charge we must account for the transfer of charge
between modes.

We have considered a (1+1)-dimensional system. Although this does
not allow us to model a condensation process, we are able to
examine the charge and energy distributions as they evolve in
time. For a sudden drop in temperature modelled by a sudden change
in the effective mass squared, we observe an expected general
movement of charge from higher to lower momentum modes. This is
accompanied by a movement of energy from lower to higher momentum
modes.

A (3+1)-dimensional version of this calculation remains a
desirable goal for the future. An improvement of the algorithm in
order to run the calculation to much later times would also be
welcome.


\section*{Acknowledgements}
This work was financially supported by The Royal Commission for
the Exhibition of 1851. I would like to acknowledge Tim Evans and
Mark Hindmarsh for illuminating discussions. I would further like
to acknowledge the hospitality of the University of Sussex and
University College London.

\end{document}